\documentstyle[12pt]{article}
\def\beq{\begin{equation}} 
\def\eeq{\end{equation}} 
\def\bea{\begin{eqnarray}}  
\def\eea{\end{eqnarray}}  
\def\bq{\begin{quote}}  
\def\eq{\end{quote}}  
  
\def\beqa{\begin{eqnarray}}  
\def\eeqa{\end{eqnarray}}  
\def\be{\begin{equation}}  
\def\ee{\end{equation}}  
\def\beq{\begin{equation}}  
\def\eeq{\end{equation}}

\def\pa{\partial}  
\def\kaps{{\kappa}^{2}}

\def\cp{{\cal P}}  
\def\cl{{\cal L}} 
\def\cf{{\cal F}}

\def\r2{\sqrt{2}}  
\def\ra{\rightarrow}
\def\bi{\begin{itemize}}  
\def\ei{\end{itemize}}

\def\ov{\overline}  
\def\nn{\nonumber \\}

\def\lc{{\cal L}}  
\def\ca{{\cal A}}

\begin{document} 

\pagestyle{empty}
\begin{flushright}   hep-th/0009167
\end{flushright}
\vskip 2cm
\begin{center}
{ \Large  \bf Five-Dimensional Gauged 
Supergravities with Universal Hypermultiplet and Warped Brane Worlds} 
\vspace*{5mm} \vspace*{1cm} 
\end{center}
\vspace*{5mm} \noindent
\vskip 0.5cm
\centerline{\large Adam Falkowski${}^1$ Zygmunt Lalak${}^{1,2}$
and Stefan Pokorski${}^{1}$}
\vskip 1cm
\centerline{\em ${}^{1}$Institute of Theoretical Physics}
\centerline{\em University of Warsaw, Poland}
\vskip 0.3cm
\centerline{\em ${}^{2}$Physikalisches Institut, Universit\"at Bonn}
\centerline{\em Nussallee 12, D-53115 Bonn, Germany}
\vskip2cm

\centerline{\bf Abstract}
\vskip 0.3cm
\noindent We present five dimensional gauged supergravity with universal hypermultiplet
on $M_4 \times S^1 / Z_2$ coupled supersymmetrically to three-branes located
at the fixed points. The construction is extended to the smooth picture with
auxiliary singlet and four-form fields. 
The model admits the Randall-Sundrum solution as a BPS vacuum with vanishing 
energy. We give the form of all KK-tower modes for fields present in the 
model.  
\vskip 6.0cm

\newpage
\pagestyle{plain}
Recently it has been realized that brane models with warped geometries 
may play a role in understanding the variety of field theoretical 
incarnations of the hierarchy problem \cite{rs1,rs2,rs3}. 
However, there are good reasons 
to believe  that supersymmetry is still an important ingredient of this 
understanding, and the quest for consistent supersymmetric versions of
brane worlds goes on, see \cite{bagger,gp,hiszp,flp,kallosh,bc,bc2,kl,kl2,kb,
pmayr,duff,louis,zucker}. The interest in 
supersymmetry follows 
from the expectation that a) it explains the fine-tuning between the bulk
cosmological constant and brane tensions necessary in such solutions, 
b) supersymmetric BPS vacua are stable against dynamical fluctuations,
hence c) 
supersymmetry may help to solve the problem of stabilization of the fifth 
dimension, d) supergravity is likely to be necessary to embed brane worlds 
in string theory.  \\
In earlier papers \cite{bc,bc2,kl,kl2} the attention has been focused on models with 
solitonic (thick) branes, however, with no positive results (but 
see \cite{pmayr,duff}). 
Finally, in papers  \cite{bagger,flp,kallosh} explicit supersymmetric models with 
delta-type (thin) branes have been constructed. 

The distinguished feature of the pure supergravity Lagrangians 
proposed in \cite{flp} is the particular way of imposing the $Z_2$ symmetry,
such that gravitino masses are $Z_2$-odd. An elegant formulation of the 
model is given in ref. \cite{kallosh} where an additional, supersymmetry 
singlet, field is introduced to independently supersymmetrize the branes 
and the bulk. In the on-shell picture for this singlet field the models 
of ref. \cite{kallosh} and ref. \cite{flp} are the same. 

In the reference \cite{kallosh} an extension of the model to include vector 
multiplets has been worked out. On the other hand, in ref. \cite{flp} it has 
been noted, that supersymmetric Randall-Sundrum-type models can be generalized 
to include the universal hypermultiplet and matter on the branes. The 
Lagrangian of such a construction, to the lowest order in fluctuations of the 
$Z_2$-odd field $\xi$ from the hypermultiplet, have been given in \cite{flp}. 

In the present paper we extend the previous results for the model with 
the universal hypermultiplet in several directions. First, we explicitly 
show how to interpret our construction in the language of gauged supergravity 
and identify the sigma-model symmetry that has to be gauged. Next, we give 
the model with the universal hypermultiplet the `smooth' formulation 
analogous to that of ref. \cite{kallosh}. In addition, we consider the 
case where vector multiplets and the universal hypermultiplet are 
simultanously present. Finally, we solve the equations of motion for 
gravitini and hyperini to find out their zero modes and the higher Kaluza-Klein
modes, to gather information about the spectrum of the low-energy effective 
theory. We shall demonstrate the fermion-boson symmetry at the level of the 
mass spectrum at all KK levels. 

We begin by recalling that supersymmetric RS Lagrangian \cite{flp} is 5d 
gauged supergravity: the gauged group is a $U(1)$ subgroup of the R-symmetry 
$SU(2)$ group, with the $SU(2)$ valued prepotential $\cp$ chosen along the 
$\sigma^3$ direction  
\beq
\label{su2prepotential}
g\cp=\frac{1}{4\r2}\Lambda \epsilon(x^5) i\sigma^3
\eeq
where $\epsilon(x^5)$ is the antisymmetric step function, $g$ is the $U(1)_R$ gauge coupling constant, and constant parameter $\Lambda$ together with 
numerical factor
are  chosen in this form for later convenience. 
Indeed, we start with the five-dimensional $N=2$ supergravity 
on $M_4 \times S^1$ with the kinetic action 
\beqa 
\label{5daction} 
&S=\int d^5xe_{5}\frac{1}{\kaps} \left ( \right .  
\frac{1}{2}R
-\frac{3}{4}{\cal F}_{\alpha\beta}{\cal F}^{\alpha\beta}
-\frac{1}{2\sqrt{2}}\epsilon^{\alpha\beta\gamma\delta\epsilon}{\cal A}_{\alpha}{\cal F}_{\beta\gamma}{\cal F}_{\delta\epsilon} -\frac{1}{2}\ov{\psi_\alpha}^A\gamma^{\alpha\beta\gamma}D_\beta\psi_{A\gamma} 
   \left . \right).  & 
\eeqa 
The model includes a gravity multiplet $\{e_\alpha^m, \psi_\alpha^A, \ca_\alpha \}$. Then we pass on to the orbifold $S^1 / Z_2$  
by moding the $S^1$ by ${\bf Z}_2$ and choosing the action of ${\bf Z}_2$ on the fields. 
Two parallel 3-branes are located at the fixed points $x^5=0$ and $x^5=\pi\rho$.

The conventions and normalizations we use are mainly those of reference \cite{ovruthet}. The signature of the metric tensor is $(-++++)$. The SU(2) spinor indices are raised with an antisymmetric tensor $\epsilon^{AB}$.
We choose $\epsilon^{12}=\epsilon_{12}=1$. The rule for dealing with symplectic spinors is $\ov{\psi_1}^A\psi_2^B=\ov{\psi_2}^B\psi_1^A$ 
(note that $\ov{\psi}^A=\ov{\psi_A}$). The  
${\bf Z}_2$ symmetry acts as  reflection $x^5 \rightarrow -x^5$ and is represented in such a way  that bosonic fields $(e_{\mu}^{m},e_{5}^{5}, {\cal A}_{5})$ are even, and $(e_{5}^{m},e_{\mu}^{5}, {\cal A}_{\mu})$ are odd. The indices $\alpha, \beta ...$  are  five dimensional ($0..3, 5$), while 4d indices are denoted by $\mu, \nu, ...$. The action of the ${\bf{Z_{2}}}$ on fermionic fields and on the parameter $\epsilon$ of supersymmetry transformations 
is defined as: 
\bea 
\label{z2} 
\gamma_{5}\psi_{\mu}^{A}(x^5)=(\sigma^{3})^{A}\;_{B}\psi_{\mu}^{B}(-x^5)& 
\gamma_{5}\psi_{5}^{A}(x^5)=-(\sigma^{3})^{A}\;_{B}\psi_{5}^{B}(-x^5) 
\nn  
\gamma_{5}\epsilon^{A}(x^5)=(\sigma^{3})^{A}\;_{B}\epsilon^{B}(-x^5)& 
\eea 
where $\gamma_{5}=(^{-{\bf 1}\;0}_{\;\;\;0\;{\bf 1}}), 
\sigma^{3}=(^{1\;\;\;0}_{0\; -1}), \; A=1,2$. Symplectic Majorana spinors  
in 5d satisfy $\bar{\chi}^A = (\chi^A)^T C$ with $C=i \gamma^2 \gamma^0\gamma^5$  
in 4d chiral representation. The supersymmetry transformations laws are
\beqa 
\label{zuzanna}
&\delta e_{\alpha}^{m}= 
\frac{1}{2}\ov{\epsilon}^A\gamma^{m}\psi_{A\alpha}& 
 \nonumber \\ 
&\delta \psi_\alpha^{A}= 
D_{\alpha}\epsilon^{A} 
-\frac{i}{4\sqrt{2}}(\gamma_{\alpha}^{\;\beta\gamma}-4\delta_{\alpha}^{\beta}\gamma^{\gamma}){\cal F}_{\beta\gamma}\epsilon^{A}
& 
\nonumber \\ 
&\delta{\cal A}_{\alpha}= 
-\frac{i}{2\sqrt{2}}\ov{\psi_\alpha}^A\epsilon_A .&  
\eeqa
The gauging of the $U(1)_R$ symmetry means 
that we replace in (\ref{5daction})  the derivative acting on the gravitino with the $U(1)_R$ covariant derivative:
\beq
D_\alpha \psi_\beta^A \rightarrow D_\alpha \psi_\beta^A 
-\frac{1}{2\r2}(\sigma^3)^A_{\;B}\Lambda\epsilon (x^5)\ca_\alpha \psi_\beta^B
\eeq 
 where $D_\alpha$ denotes the ordinary space-time covariant derivative. The global U(1) symmetry $\psi_\alpha^A \rightarrow e^{i\phi}\psi_\alpha^A$ of the pure 5d supergravity has been promoted to a local symmetry, the gauge field being 
$\ca_\alpha^R=-\frac{1}{2\r2}\ca_\alpha$. In the same way we modify the derivative acting on the supersymmetry parameter $\epsilon^A$ (it also carries an SU(2) index) in the SUSY transformation laws.
The gauging produces various new terms in the bulk Lagrangian:
\beq 
\label{gmass} 
\lc_{\psi^2}=+\frac{e_5}{8\kaps}\Lambda\epsilon(x^5) 
( 
\ov{\psi_\alpha^1}\gamma^{\alpha\beta}\psi_\beta^1 
-\ov{\psi_\alpha^2}\gamma^{\alpha\beta}\psi_\beta^2 
), 
\eeq 
\beq 
\lc_{A}= 
-\frac{ie_5}{4\sqrt{2}\kaps}\epsilon(x^5)\Lambda 
\left ((\ov{\psi^1}_\alpha\gamma^{\alpha\beta\gamma}\psi^1_\gamma){\cal A}_\beta 
 -(1\rightarrow 2)\right ) \eeq 
and a bulk cosmological term
\beq 
\label{cosmo} 
\lc_{C}=\frac{e_5}{6\kaps}\Lambda^2. 
\eeq 
The presence of the $\epsilon (x^5)$ in the prepotential $\cp$ requires 
branes with compensating sources
\beq
V_1= -V_2 = \Lambda.
\eeq
In the original papers of Randall and Sundrum this apparent fine-tuning was not justified,
whereas in the present construction it follows from the requirement of 
supersymmetry of the complete, bulk and branes Lagrangian.\\
This relation  is 
exactly the one we need to get the Randall-Sundrum solution with the warp 
factor $a=exp(-\frac{\Lambda b_0}{6} |x^5|)$ (the brane world metric is 
$ds^2 = a^2(x^5) \eta_{\mu\nu} dx^\mu dx^\nu + b_0^2 (dx^5)^2$). In the notation used in \cite{rs1,rs2} $k=\Lambda/6$.

In the ordinary 5d supergravity, gaugings of different U(1) subgroups of  the SU(2) R-symmetry group (in other words, choices of a direction of the prepotential in the space of $SU(2)$ generators) can be rotated into each other. However in the present  framework, the orbifold ${\bf Z}_2$ symmetry does not commute with the SU(2) R-symmetry and gauging different $U(1)_R$ results in different theories.
E.g. the authors of \cite{bagger} choose $\cp \sim \Lambda \sigma^2$ and obtain a different form of the boundary Lagrangian.
We stress again, that it is crucial for our construction that the U(1) gauge coupling constant 
changes sign when it crosses the brane.
This completes the previous construction of \cite{flp}.

In the construction of ref. \cite{kallosh} the odd gauge charge is replaced by a new field $G$ which is 
invariant under supersymmetry transformations 
\beq
\Lambda \epsilon(x^5) \ra G(x^\alpha)
\eeq
With this replacement the bulk action is no longer supersymmetric. 
The SUSY variation of the bulk Lagrangian gives:
\bea
\label{var1}
\delta\cl= e_5\frac{i}{4} \left(i\ov{\psi_\alpha}^A \gamma^{\alpha\beta}\epsilon^B \sigma^{3}_{AB}\pa_\beta G
-\r2\ov{\psi_\alpha}^A \gamma^{\alpha\beta\gamma}\epsilon^B \sigma^{3}_{AB}
 {\cal A}_\beta \pa_\gamma G  \right ).
\eea
To cancel the above variation the authors of \cite{kallosh} propose to 
introduce a new four-form field
 $G_{\alpha\beta\gamma\delta}$. A  new term is added to the Lagrangian which couples the four-form to the scalar:
\beq
\label{var2}
\cl=\frac{1}{4!}\epsilon^{\alpha\beta\gamma\delta\epsilon}G_{\alpha\beta\gamma\delta}\pa_\epsilon G
\eeq
Then the  SUSY transformation of the four-form G is defined in such  way 
that it  cancels (\ref{var1}):
\beq
\frac{1}{4!}\epsilon^{\alpha\beta\gamma\delta} \delta G_{\alpha\beta\gamma\delta}=
e_5 \frac{i}{4} \left(i\ov{\psi_\alpha}^A \gamma^{\alpha\epsilon}\epsilon^B 
\sigma^{3}_{AB}
-\r2\ov{\psi_\alpha}^A \gamma^{\alpha\beta\epsilon}\epsilon^B \sigma^{3}_{AB}
{\cal A}_\beta \right ).
\eeq
To make contact with the previous scheme with odd gauge charges one adds boundary terms to the Lagrangian:
\beq
\label{var3}
\cl_B =  -\Lambda\left (
\frac{1}{4!}\epsilon^{\alpha\beta\gamma\delta} G_{\alpha\beta\gamma\delta}
+e_4 \right ) (\delta(x^5) - \delta(x^5-\pi\rho))
\eeq
The SUSY variation of the four-form $G$ cancels against the SUSY variation of the second term up to the expression containing only odd fields (which vanish on the branes).   
Varying the action with respect to $G_{\alpha\beta\gamma\delta}$ yields the equation of motion for the scalar G:
\beq
\pa_5 G = 2\Lambda (\delta(x^5) - \delta(x^5-\pi\rho))
\eeq
which is solved by $G=\Lambda \epsilon(x^5)$. Substituting this solution 
 back to the Lagrangian we find again the Lagrangian of gauged supergravity 
with odd gauge charges. Thus upon integrating out the field $G$ both schemes 
are equivalent. The formulation using the four-form $G$ has the virtue that 
all commutators of the superalgebra are easily seen to close.

Now we go on and expand the pure supergravity model by the addition 
of a universal hypermultiplet. Such a multiplet appears in C-Y 
type compactifications of string and M-theory Lagrangians down to five 
dimensions. The universal hypermultiplet $(\lambda^a, V, \sigma, \xi, \bar{\xi})$ forms a  $SU(2,1)/U(2)$ non-linear sigma model. 
The metric of this sigma-model, which is also a quaternionic manifold,
can be read from the K\"ahler potential: 
$ K=-ln(S+\bar{S}-2\xi\bar{\xi}), \;
S=V+\xi\bar{\xi}+i\sigma. $ 
The new terms in the kinetic part of the bulk $N=2$ Lagrangian are given by
\beqa 
\label{5dactionhip} 
&S=\int d^5xe_{5}\frac{1}{\kaps} \left ( \right .   
-\frac{1}{4V^2}(\pa_{\alpha}V\pa^{\alpha}V+\pa_{\alpha}\sigma \pa^{\alpha}\sigma) 
& \nonumber \\
&-\frac {1}{V}\pa_{\alpha}\xi\pa^{\alpha}\bar{\xi}
-\frac{i}{4V^2}(\xi\pa_{\alpha}\bar{\xi} \pa^{\alpha}\sigma-\bar{\xi}\pa_{\alpha}\xi \pa^{\alpha}\sigma)
+\frac{1}{4V^2}( 
(\xi\pa_{\alpha}\bar{\xi})^2+(\bar{\xi}\pa_{\alpha}\xi)^2 
-|\bar{\xi}\pa_{\alpha}\xi|^2) & \nonumber \\ 
&  -\frac{1}{2}\ov{\lambda}^a\gamma^\alpha D_\alpha \lambda_a \left . \right).  & 
\eeqa 
The Sp(1) indices on the hyperini are raised and lowered with $\Omega^{ab}$.
We choose $\Omega^{21}=\Omega_{21}=1$.
The ${\bf Z}_2$ symmetry is represented in such a way  that bosonic fields 
$(V, \sigma)$ are even, and $(\xi)$ is odd. 
The action of the ${\bf{Z_{2}}}$ on hyperini
is given by: 
\beq
\label{zh} 
\gamma_{5}\lambda^{a}(x^5)=-(\sigma^{3})^{a}\;_{b}\lambda^{b}(-x^5). 
\eeq
The ungauged Lagrangian with hypermultiplet is supersymmetric with the 
following transformations of the fields
\beqa 
&\delta V= 
\frac{i}{\sqrt{2}}V(\ov{\epsilon^1}\lambda^1)-(1\rightarrow 2)& 
\nonumber \\ 
&\delta \sigma = 
+\frac{1}{\sqrt{2}}V(\ov{\epsilon^1}\lambda^1)+(1\rightarrow 2) 
+\sqrt{\frac{V}{2}}(\xi\ov{\epsilon^1}\lambda^2-\bar{\xi}\ov{\epsilon^2}\lambda^1)& 
\nonumber \\ 
&\delta \xi= 
-\frac{i\sqrt{V}}{\sqrt{2}}(\ov{\epsilon^2}\lambda^1) 
\; \; \;  
\delta \bar{\xi}= 
-\frac{i\sqrt{V}}{\sqrt{2}}(\ov{\epsilon^1}\lambda^2)& 
\nonumber \\ 
&\delta \lambda^1= 
-\frac{i}{2\sqrt{2}V} (\pa\!\!\!\slash(V+i\sigma) 
-\bar{\xi}\pa\!\!\!\slash\xi+\xi\pa\!\!\!\slash\bar{\xi} ) \epsilon^1 
+\frac{i}{\sqrt{2V}}\pa\!\!\!\slash\xi\epsilon^2& 
\nonumber \\ 
&\delta \lambda^2= 
+\frac{i}{2\sqrt{2}V} ( \pa\!\!\!\slash(V-i\sigma) 
+\bar{\xi}\pa\!\!\!\slash\xi-\xi\pa\!\!\!\slash\bar{\xi} ) \epsilon^2 
+\frac{i}{\sqrt{2V}}\pa\!\!\!\slash\bar{\xi}\epsilon^1 \, .& 
\eeqa  
If  the universal hypermultiplet is present in the gauged 5d action we need 
further modifications to arrive at  a supersymmetric Lagrangian. In \cite{flp} it was found that to cancel variations of the form $\sim \Lambda \epsilon \lambda \pa_\alpha V$ it is necessary to introduce a hyperino mass term:
\beq 
\label{hyperinomass}
\cl_{\lambda^2}= 
\frac{e_5}{8\kaps}\epsilon(x^5)\Lambda \left (\ov{\lambda^1}\lambda^1 
-(1\rightarrow 2)\right )
\eeq 
The presence of the hyperino  mass terms in the Lagrangian indicates that to obtain a supersymmetric theory it is necessary to gauge some isometry of the hypermultiplet scalar manifold. 
Indeed, in the general 5d supergravity Lagrangian coupled to hypermultiplets the mass term of the hypermultiplet fermion has the form 
$
 -\frac{ig}{4\r2}V_u^{Aa}V_v^{Bb}\epsilon_{AB} \nabla^{[u}k^{v]} \ov{\lambda}_a \lambda_b
$
where $V_u^{Aa}$ is the vielbein describing the sigma model metric and $k^u$ is a Killing vector which generates a symmetry transformation on the scalar manifold:
$q^u \ra q^u + k^u$.
The hyperino mass term  is present only when the Killing vector $k^u$ of the gauged isometry is non-zero.

Recall \cite{gunaydin,agata} that in the case of quaternionic manifolds  a  Killing vector can be determined by a Killing prepotential according to the formula:
\beq
\label{killingprepotential}
k^u K_{uw} = \pa_{w} \cp + [\omega_w,\cp] 
\eeq
where  $K_{uv}$ is the K\"ahler form and  $\omega_v$ the spin connection of the quaternionic  manifold and $\cp$ is the Killing prepotential. In the standard formulation of gauged 5d supergravity \cite{agata} the Killing prepotential is equal to the SU(2) prepotential which determines gauging of the SU(2) R-symmetry \footnote{However, authors of \cite{louis} argue that this condition can be relaxed.}.

The scalar potential can be calculated using the formula:
\beq
\label{scalarpotential}
V= g^2(\frac{8}{3} tr (P^2)+\frac{1}{2}h_{uw}k^u k^w)
\eeq 
 where $h_{uw}$ is the metric of the hypermultiplet's scalar manifold. To remain in the framework of the Randall and Sundrum model we have to find a Killing vector leading to a scalar potential which is negative at its critical point, so that we can obtain $AdS_5$ solutions.  

There are the following isometries in our model:
\beq \label{first}
\xi \rightarrow \xi +i\beta  \hspace{1cm} S \rightarrow S - 2i \beta \xi + 
\beta^2
\eeq
\beq \label{second}
V \rightarrow V + \gamma Re \xi \hspace{1cm} \sigma \rightarrow \gamma Im \xi
\eeq
\beq \label{third}
\sigma \rightarrow \sigma + \alpha  
\eeq
\beq \label{fourth}
V \rightarrow \kappa V \hspace{1cm} \sigma \rightarrow \kappa \sigma \hspace{1cm}
\xi \rightarrow \kappa^{1/2}\xi
\eeq
\beq \label{fifth}
\xi \rightarrow e^{i\theta} \xi
\eeq
The first (\ref{first}) and the second (\ref{second}) isometries imply 
that their gauge coupling constants $g$ are even. This can be easily seen from the form of  covariant derivatives $D_\mu q^u= \pa_\mu q^u + g k^u \ca_\mu$ . For example, for the case of the first isometry (\ref{first}) the covariant derivative acting on $\xi$ is $\pa_\mu \xi + g \ca_\mu$. The 
$\pa_\mu \xi$ is odd and since  components of the graviphoton $\ca_\mu$ along the four dimensions are odd, the gauge coupling constant must be even for the whole expression to have definite ${\bf Z}_2$ parity. Even gauge coupling 
constant does not lead to discontinuities in the SUSY transformation laws. As explained in previous paragraphs, it is the appearance of the ${\bf Z}_2$ odd step function $\epsilon(x^5)$ in the transformation laws that allows us to introduce brane potentials in a supersymmetric way.

The remaining isometries (\ref{third},\ref{fourth},\ref{fifth}) lead to ${\bf Z}_2$ odd coupling constants. 
The third one, the 
shift of the axion (\ref{third}), has been studied in \cite{ovrutdw,ovruthet,elp}. 
It leads to the potential of the form $\frac{\alpha^2}{V^2}$ which does not admit $AdS_5$ vacuum solution. It can be shown that also the fourth of the given isometries, the scale transformations (\ref{fourth}), leads to a scalar potential which does not admit $AdS_5$ solutions.

We will show that the last of the isometries, (\ref{fifth}), leads to a scalar potential of the desired form. This isometry is present only in the case of the 
universal hypermultiplet. Thus it is not straightforward to extend our construction to the non-universal case.

To perform calculations of the prepotential it is convienient to work with real coordinates, hence we define $\xi= x + i y$. In these coordinates the components of the spin connection and the K\"ahler form can be expressed in terms of the Pauli matrices:
\bea
\omega_V=0  
&
\omega_\sigma=\frac{1}{4V} i \sigma^3
\nn
\omega_x= \frac{y}{2V} i\sigma^3 - \frac{1}{V^{1/2}}i\sigma^2
&
\omega_y= -\frac{x}{2V} i\sigma^3 - \frac{1}{V^{1/2}}i\sigma^1
\eea
\bea
K_{V\sigma}=\frac{1}{8V^2}i\sigma^3
&
K_{Vx}=-\frac{1}{4V^{3/2}}i\sigma^2 + \frac{y}{4V^2}i\sigma^3
&
K_{Vy}=-\frac{1}{4V^{3/2}}i\sigma^1 - \frac{x}{4V^2}i\sigma^3
\nn
K_{\sigma x}=\frac{1}{4V^{3/2}}i \sigma^1
&
K_{\sigma y}=-\frac{1}{4V^{3/2}}i \sigma^2
&
K_{xy}=-\frac{1}{2V} i \sigma^3 +  \frac{x}{2V^{3/2}}i\sigma^1 -\frac{y}{2V^{3/2}}i\sigma^2
\eea
The non-vanishing components of the Killing vector corresponding to the studied isometry are:
\beq
\label{ourkilling}
k^x=-y \;\;\; k^y=x .
\eeq
Since $\omega_V=0$, the $w=V$ component of equation (\ref{killingprepotential}) determines $\cp$ up to an arbitrary $V$-independent matrix:
\beq
\cp=- \frac{x^2+y^2}{4V} i \sigma^3 - \frac{x}{2V^{1/2}} i\sigma^1 
+ \frac{y}{2V^{1/2}} i\sigma^2 +f(x,y,\sigma).
\eeq
The $w=\sigma$ component hints that the matrix $f$ is proportional to $\sigma^3$ and independent of the field $\sigma$. Finally, the $w=x$ component implies that the matrix $f$ is a constant matrix and the final form of the prepotential is:
\beq
\label{ourprepotential}
g\cp= g \left ( 
(\frac{1}{4}- \frac{x^2+y^2}{4V}) i \sigma^3 - \frac{x}{2V^{1/2}} i\sigma^1 
+ \frac{y}{2V^{1/2}} i\sigma^2 ) 
\right ).
\eeq
To make contact with our previous notation we define $g=\Lambda\epsilon(x^5)/\r2$. With this substitution the constant part of the obtained prepotential is the same as in (\ref{su2prepotential}). Compared to the case of pure supergravity in the bulk, we see that $\xi$ dependent term appears in the prepotential.
It is straightforward to check that with the Killing vector (\ref{ourkilling}) and the prepotential (\ref{ourprepotential}) we derive the same hyperino masses as in (\ref{hyperinomass}) up to $\xi$-dependent corrections.

Using (\ref{scalarpotential}) we calculate the scalar potential:
\beq
V = - \Lambda^2 \left ( \frac{1}{6}
 +\frac{1}{12 V}|\xi|^2 - \frac{1}{12V^2}|\xi|^4 \right).
\eeq   
As before, the presence of the step function $\epsilon(x^5)$ in the prepotential leads to  discontinuity in the SUSY transformation law. In consequence, variation of  the bulk Lagrangian contains terms proportional to the delta functions at the boundaries. These can be cancelled by adding the brane potential to the Lagrangian.  The SUSY transformation law of the gravitino contains a piece:
\beq
\delta\psi_\alpha^A=\frac{1}{12}\Lambda \epsilon(x^5) (1-\frac{|\xi|^2}{V})
 (\sigma^3)^A_{\;B}\epsilon^B .
\eeq 
As a consequence, the variation of the gravitino kinetic term contains a 
singular expression:
\bea
&\delta \cl=
\frac{e_5}{\kaps_5}(\ov{\psi_\mu^1}\gamma^{\mu \nu}\gamma^5\pa_5(\frac{\Lambda}{12}\epsilon(x^5)\gamma_\nu\epsilon^1) (1-\frac{|\xi|^2}{V})
-(1 \rightarrow 2)
&\nn&
=-\frac{e_5}{2\kaps_5}\Lambda\delta(x^5)(\ov{\psi_\mu^1}\gamma^\mu\epsilon^1) (1-\frac{|\xi|^2}{V})
+(1 \rightarrow 2)+\dots
&\eea
This can be cancelled by adding a potential term localized on  the branes:
\beq 
\label{brane1} 
\lc_B=-\frac{e_4}{\kaps}\Lambda  (1-\frac{|\xi|^2}{V}) (\delta(x^5) - \delta(x^5-\pi\rho)).  
\eeq   
This modification is also consistent with the new corrections to hyperino 
transformation laws
$
\delta \lambda^a = \frac{\Lambda \epsilon(x^5)}{2}V_u^{Aa}k^u \epsilon_A
$.

The  model we consider in this paper can be formulated in a more elegant way, by replacing the odd gauge coupling constant with a scalar field, in the same way as it was done in \cite{kallosh}:
\beq
\Lambda \epsilon(x^5) \ra G(x^\alpha) .
\eeq
With this replacement the bulk action ceases to be supersymmetric. Following (\ref{var1}) we get:
\bea
\label{varh1}
\delta\cl= e_5\left(i\ov{\psi_\alpha}^A \gamma^{\alpha\beta}\epsilon^B \cp_{AB}\pa_\beta G
-\r2\ov{\psi_\alpha}^A \gamma^{\alpha\beta\gamma}\epsilon^B \cp_{AB}
{\cal A}_\beta \pa_\gamma G
+\frac{1}{2} \ov{\lambda}_a\gamma^\alpha \epsilon_A V_u^{Aa} k^u \pa_\alpha G    \right )
\eea
which can be cancelled by modifying the action and the transformation laws analogously to (\ref{var2}), (\ref{var3}):
\beq
\cl=\frac{1}{4!}\epsilon^{\alpha\beta\gamma\delta\epsilon}G_{\alpha\beta\gamma\delta}\pa_\epsilon G,
\eeq
\beq
\label{kalloshbrane}
\cl_B =  -\Lambda\left (
\frac{1}{4!}\epsilon^{\alpha\beta\gamma\delta} G_{\alpha\beta\gamma\delta}
+e_4(1-|\xi|^2) \right ) (\delta(x^5) - \delta(x^5-\pi\rho))
\eeq
\beq \label{cvarh1}
\frac{1}{4!}\epsilon^{\alpha\beta\gamma\delta} \delta G_{\alpha\beta\gamma\delta}=
e_5\left(i\ov{\psi_\alpha}^A \gamma^{\alpha\epsilon}\epsilon^B \cp_{AB}
-\r2\ov{\psi_\alpha}^A \gamma^{\alpha\beta\epsilon}\epsilon^B \cp_{AB} {\cal A}_\beta 
+\frac{1}{2} \ov{\lambda}_a\gamma^\epsilon \epsilon_A V_u^{Aa} k^u     \right ).
\eeq
The SUSY variation of the four-form $G$ in (\ref{kalloshbrane}) cancels against the SUSY variation of the second term up to the expression containing only odd fields (which vanish on the branes).   
Varying the action with respect to $G_{\alpha\beta\gamma\delta}$ yields the equation of motion for the scalar G:
\beq
\pa_5 G = 2\Lambda (\delta(x^5) - \delta(x^5-\pi\rho))
\eeq
which has the solution $G=\Lambda \epsilon(x^5)$. After putting it back to the Lagrangian we obtain, again, the Lagrangian of gauged supergravity with odd gauge charges. Thus after integrating out the additional fields both schemes are equivalent also in the presence of the hypermultiplet. 
In spite of the presence of additional $\xi$ dependent terms in the bulk and boundary potential, the $AdS_5$ metric  is still a solution to the equations of motion. The equation of motion for the scalar $\xi$ is:
\beq
\label{ksi}
\xi'' +4\frac{a'}{a}\xi'+\frac{(\Lambda b_0)^2}{12}\xi = -\xi(\delta(x^5)-\delta(x^5-\pi\rho))
\eeq
where we assumed that $V'=\sigma{}'=0$.
It is clear that $\xi=0$ is a perfectly good solution to the above equation.  With $\xi=0$ the scalar potential reduces to $V= \frac{1}{6} \Lambda ^2 + \Lambda (\delta(x^5)- \delta(x^5-\pi\rho))$  and the Einstein equations yield indeed the Randall-Sundrum solution for the warp factor $a(x^5)$. 

Having obtained the complete supersymmetric extension of the 
Randall-Sundrum scenario in the presence of the matter hypermultiplet in the bulk we can now try to generalize the model by gauging the other isometries we have at hand. Particularly interesting is to gauge the axion shift $ \sigma \ra \sigma + \alpha$ because this leads to couplings in the Lagrangian which occur in the M-theory compactified on the Calabi-Yau manifold \cite{ovrutdw}. The Killing vector corresponding to this symmetry is $k^\sigma = -2$ and the Killing prepotential determined by the equation (\ref{killingprepotential}) is  $\cp = -\frac{1}{4V}i\sigma^3$. More generally we can consider the Killing vector to be a linear combination of the two symmetries:
\beq \label{prepab}
k^x=-\beta y \;\;\; k^y = \beta x \;\;\; k^\sigma = -2 \alpha 
\eeq
where $\alpha$, $\beta$ are arbitrary real numbers. The Killing prepotential for such isometry is:
\beq \label{kilab}
g\cp= g \left ( 
(\beta\frac{1}{4}-\alpha\frac{1}{V}- \beta\frac{x^2+y^2}{4V}) i \sigma^3 - \beta\frac{x}{2V^{1/2}} i\sigma^1 
+ \beta\frac{y}{2V^{1/2}} i\sigma^2 ) 
\right ) .
\eeq
Choosing $g =\frac{\Lambda\epsilon(x^5)}{\r2}$ the scalar potential is:
\beq
V=\Lambda^2 \left (
-\frac{1}{6}(\beta-\alpha\frac{1}{V})^2
+\alpha^2 \frac{1}{4V^2}
-\beta^2\frac{1}{12V}|\xi|^2
-\alpha\beta\frac{5}{6V^2}|\xi|^2
+\beta^2\frac{1}{12V^2}|\xi|^4 
             \right ) .
\eeq
Gauging of the axion shift introduces also a new term to the brane potential
\beq 
\label{brane2} 
\lc_B=-\frac{e_4}{\kaps}\Lambda  (\beta-\alpha\frac{1}{V}-\beta\frac{|\xi|^2}{V}) (\delta(x^5) - \delta(x^5-\pi\rho)).  
\eeq  
 
In the present scheme it is straightforward to introduce vector multiplets in the bulk, in the same way as it was done in \cite{kallosh}. In the notation of \cite{ovruthet} one adds a correction to the prepotential $\cp$ which depends
on the scalars $\phi^x$ belonging to vector multiplets. Such correction 
has to respect the equation (\ref{killingprepotential}) for given Killing vectors
and given geometry of the quaternionic manifold. In case of the universal 
hypermultiplet and Killing vector (\ref{kilab}) the prepotential (\ref{prepab})is  shifted in the way $\cp \rightarrow \cp + \cp_V$ where 
$\cp_V= g(\phi^x) i \sigma^3$ and $g(\phi^x)$ is a function of 
scalars from the vector multiplets. The resulting correction to 
(\ref{varh1}) is 
\beqa
\label{varh2}
&\delta\cl= e_5\left(i\ov{\psi_\alpha}^A \gamma^{\alpha\beta}\epsilon^B \cp_{AB}\pa_\beta G
-\r2\ov{\psi_\alpha}^A \gamma^{\alpha\beta\gamma}\epsilon^B \cp_{AB}
{\cal A}_\beta \pa_\gamma G - \frac{1}{2} \bar{\lambda}^{A}_{x} 
\cp_{V\;AB}^{'x} \gamma^\alpha \partial_\alpha G \right . & \nonumber \\
&+ \left . \frac{1}{2} \ov{\lambda}_a\gamma^\alpha \epsilon_A V_u^{Aa} k^u \pa_\alpha G    \right )&
\eeqa
and respective corrections needed in (\ref{cvarh1}) and in (\ref{brane2}) 
immediately follow. In the above $\lambda^{A}_{x}$ denotes 
symplectic-Majorana fermions from vector multiplets $\{A_\alpha, \; 
\phi^x, \;\lambda^{A}_{x} \}$  which are labelled by the 
index $x$, and the apostrophe means the differentiation with respect to $\phi^x$: $ \cp^{'x} = g^{xy} \partial \cp / \partial \phi^y$ ($g^{xy}$ being the sigma model metric in the vector multiplet sector). 

{}Finally, it is immediate to include into the scheme in the supersymmetric way gauge sectors with four-dimensional gauge and chiral matter 
supermultiplets living on the branes, which was  achieved in the reference \cite{flp,mt}. We do not repeat the details here, as all the corrections 
to the supersymmetry transformation laws and the brane-bulk couplings which are not discussed in the present note are given explicitly there, see \cite{mt} for details. \\
\vskip0.5cm
In the remaining part of the paper we turn to study the spectrum of 4d 
fluctuations of various fields. 
We work, as usual, in the `upstairs' picture, i.e. 
on the circle underlying the $S^1/Z_2$ orbifold.
The 5d theory has N=2 and the RS background preserves half of the supercharges \cite{flp, kallosh}. Thus the effective 4d theory should exhibit N=1 supersymmetry. To investigate the spectrum of the effective theory we study excitations of the 5d fields about the vacuum solution. We begin with the metric and write: 
$
g_{\mu\nu}(x^\mu,x^5) =a^2(x^5)\eta_{\mu\nu}+\phi_h (x^5) h_{\mu\nu} (x^\mu)
$
We work in the gauge 
 $\pa^\mu h_{\mu\nu}=h^\mu_\mu=0$. We also factorize out an $x^5$ dependent part $\phi_h$ and  assume that $h_{\mu\nu}$ is a mass eigenstate of the 4d d'Alambertian: $\eta^{\rho\sigma}\pa_\rho\pa_\sigma h_{\mu\nu}= m^2 h_{\mu\nu}$ . To linear order in $h$ the Einstein equations reduce to just one equation \cite{rs1}:
\beq
 -\frac{1}{2}\phi_h''+2(b_0k)^2 \phi_h
-\frac{1}{2}b_0^2 m^2 e^{-\frac{b_0\Lambda}{3}|x^5|} \phi_h
= 2b_0 k (\delta(x^5) - \delta(x^5-\pi\rho))\phi_h
\eeq
where  we defined $k=\frac{\Lambda}{6}$. When we calculate the low energy spectrum we assume for definiteness that the size of the fifth dimension is $\pi\rho \sim 37$ and call the branes Planck (at $x^5=0$) and TeV at $x^5=\pi\rho$).

{}For $m=0$ the solution is $\phi_h =A_0 e^{-2b_0k|x^5|}= a^2 (x^5)$ , so that the zero mode of the graviton is just $x^5$ independent oscillation about the vacuum solution. The wave function of the massless graviton is localized on the Plank brane at $x^5=0$.

{}For non-zero $m$ the solution is \cite{rs1}:
\beq
 \phi_h = A_m J_2 (\frac{m}{k}e^{b_0k |x^5|})
+B_m Y_2   (\frac{m}{k}e^{b_0k |x^5|}) .
\eeq
In the above formula  $J_2$ and $Y_2$ denote the Bessel and the Neumann function of the second kind, respectively. The relation between constants $A_m$ and $B_m$ can be determined by matching the delta function at $x^5=0$ in the equation motion:
\beq
\label{ab}
B_m=-A_m\frac{J_1(\frac{m}{k})}{Y_1(\frac{m}{k})} .
\eeq
Thus the solution for both the massless and the massive modes is determined up to a multiplicative constant $A_m$ which can be fixed by canonical normalization of the graviton in the 4d theory. This feature will also be present for the remaining bulk fields we consider below. 

The parameter $k$ is assumed to be of the order of the Planck scale. For the KK masses of interest the ratio $m/k$ is very small and we can safely approximate the Bessel functions with their asymptotes near zero and  the above formula reduces to
$
B_m = \frac{\pi}{4}\frac{m^2}{k^2}A_m
$

Matching delta singularities on the second brane implies the quantization of the graviton mass. Namely, $m$ must solve the equation:
\beq
\label{massq}
A_m J_1(\frac{m}{k}e^{b_0 k\pi\rho}) +B_m Y_1(\frac{m}{k} e^{b_0 k\pi\rho})=0.
\eeq
Note  for small KK masses $B_m$ is much smaller than $A_m$, so away from the brane at $x^5=0$ the contribution of the Neumann function $Y$ to the solution is negligible and the mass quantization condition is the condition for zeros of the Bessel function $J_1$ which are approximately:
$
m \approx  (n+\frac{1}{4})\pi k e^{-b_0 k \pi\rho}
$.

We now turn to the other bulk fields which are present in the supersymmetric Randall-Sundrum scenario. To solve the equation of motion for the graviphoton we 
work in the gauge 
 $\ca_5=0$. Again we factorize out the $x^5$ dependent part, so that 
$\ca_\mu (x^\mu, x^5) = \phi_A (x^5) \ca_\mu(x^\mu)$ and assume that $ \ca_\mu(x^\mu)$ satisfies the 4d Proca equation $\pa_\mu \cf^{\mu\rho}= -m^2\ca^\rho$. The equation to solve is:
\beq
\label{graviphotonequation}
\phi_A ''- 2k\epsilon(x^5) \phi_A ' +m^2 b_0^2 e^{-2k|x^5|} \phi_A=0.
\eeq
{}For $m=0$ the solution is $\phi_A= \epsilon(x^5)(M e^{2k|x^5|} +N)$, where $\epsilon(x^5)$ has  been added to ensure that $\ca_\mu$ is  ${\bf Z}_2$ odd. But matching the delta functions yields $\phi_A=0$. Indeed, non-zero $M$, $N$ would imply that the second derivative of $\phi_A$ is proportional to the derivative of the delta function, but there is nothing to cancel $\delta{}'{}(x^5)$ in the equation of motion  (\ref{graviphotonequation}). Thus, there is no massless field corresponding to the graviphoton in the effective 4d theory.
However for positive $m$ there are non-zero solutions, namely:
\beq
\phi_A = \epsilon(x^5) e^{k|x^5|} \left (
A_m J_1(\frac{m}{k}e^{k |x^5|})
+B_m Y_1   (\frac{m}{k}e^{k |x^5|})   \right ) .
\eeq
As advocated in the discussion of the zero-mode, $\phi_A$ has to vanish at the fixed points to match the delta functions. The condition $\phi_A(0)=0$ yields:
$
\frac{A_m}{B_m} = - \frac{Y_1(\frac{m}{k})}{J_1(\frac{m}{k})}
$
 which gives the same relation between the coefficients $A$ and $B$ as in the solution for the graviton (\ref{ab}). With this in mind the condition $\phi_A(\pi\rho)=0$ yields just the mass quantization (\ref{massq}). Thus the KK modes of the graviphoton on the half-circle have exactly the same mass as the KK modes of the graviton. Note also that the condition $\phi_A(0)=\phi_A(\pi\rho)=0$ implies that at every KK level the graviphoton decouples from the boundary.

Let us now study massless and massive  excitations of the gravitino $\psi^A$. 
We work in the gauge 
$\psi_5=0$.
 We factorize the $x^5$ dependence of the gravitino in the following way:
\bea
\label{gravitinoansatz}
(\psi^1_\mu)_R(x^\mu,x^5)=\phi_\psi(x^5) (\psi_\mu)_R (x^\mu)
&
(\psi^2_\mu)_L(x^\mu,x^5)=\phi_\psi(x^5) (\psi_\mu)_L (x^\mu)
 \nn
(\psi^1_\mu)_L(x^\mu,x^5)=\phi^-_\psi(x^5) (\psi_\mu)^-_L (x^\mu)
&
(\psi^2_\mu)_R(x^\mu,x^5)=\phi^-_\psi(x^5) (\psi_\mu)^-_R (x^\mu) .
\eea
To match the ${\bf Z}_2$ properties of the gravitino  $\phi_\psi(x^5)$ must be even while $\phi_\psi^-(x^5)$ must be odd. The 4d spinors  $\psi(x^\mu), \psi^-(x^\mu)$ satisfy the 4d Majorana condition.  Moreover, they are assumed to satisfy the 4d Rarita-Schwinger eqaution with mass $m$:
\bea
\gamma^{\mu\nu\rho}\pa_\nu\psi_\rho=-m\gamma^{\mu\rho}\psi_\rho
\nn
\gamma^{\mu\nu\rho}\pa_\nu\psi^-_\rho=m\gamma^{\mu\rho}\psi^-_\rho
\eea
 Plugging (\ref{gravitinoansatz}) and the RS background for the metric into the 5d equation of motion for the gravitino yields coupled equations:
\bea
\label{gravitinoequation}
\pa_5\phi_\psi+\frac{1}{2} k \epsilon(x^5) \phi_\psi - b_0 m e^{k|x^5|} \phi_\psi^- = 0
\nn
\pa_5\phi_\psi^- -\frac{5}{2} k \epsilon(x^5) \phi_\psi^- + b_0 m e^{k|x^5|} \phi_\psi  = 0 .
\eea
{}For m=0 the equations decouple. The solution for the even function $\phi_\psi$ is:
\beq
\phi_\psi=A_0 e^{-\frac{k}{2}|x^5| } \equiv A_0 a^{1/2} .
\eeq
The zero modes of the odd components of the gravitino have to vanish for exactly the same reason as vanishes the zero mode of the graviphoton; otherwise $\pa_5\phi^-$ contains the delta function which cannot be cancelled in the equation of motion (\ref{gravitinoequation}). Thus, in the effective 4d theory there is only one massless Majorana gravitino, which enters the N=1 gravity multiplet as the superpartner of the metric. This mode is localized on the Planck brane at $x^5=0$. 

{}For massive modes on the half-circle both the even and the odd combinations 
of the 5d gravitini are non-zero and the solution is:
\bea
\phi_\psi =e^{\frac{3}{2}b_0 k |x^5|}
( A_m J_2(\frac{m}{k}e^{b_0 k|x^5|}) +B_m Y_2(\frac{m}{k} e^{b_0 k|x^5|}))
\nn
\phi_\psi^-=i \epsilon(x^5)e^{\frac{3}{2}b_0 k |x^5|}(
 A_m J_1(\frac{m}{k}e^{b_0 k|x^5|}) +B_m Y_1(\frac{m}{k} e^{b_0 k|x^5|})).
\eea
Already at this point it comes as no surprise that matching delta functions yields exactly the relation (\ref{ab}) for the coefficients $A_m$, $B_m$ and the mass $m$ of the gravitini must be quantized according to (\ref{massq}). Thus at every KK level we have two Majorana gravitini, with the same mass as a heavy graviton and a  graviphoton.
Note also that the odd gravitino, just like the graviphoton, vanishes at the ${\bf Z}_2$ fixed points and thus decouples from the branes. 
        
The situation is similar in the hypermultiplet sector. Only the even components of the 5d hyperini  contribute to the massless mode. The solutions of the 
zero mode equation can be grouped into a 4d Majorana spinor $\lambda$ with components
$\lambda_R = \phi_\lambda \lambda^2_R$, $\lambda_L = \phi_\lambda \lambda^1_L$ and their profile $\phi_\lambda$ along the fifth dimension is :
\beq
\phi_\lambda = A_0 e^{\frac{1}{2}b_0 k |x^5|} \equiv    A_0 a^{-1/2}(x^5).  
\eeq
In the 4d effective theory the massless hyperino enters a N=1 multiplet together with the (constant along $x^5$) zero modes of the even hypermultiplet scalars $V$, $\sigma$. Neither hyperino nor scalars are localized on the Planck brane, which was the case for the zero modes of the gravity multiplet. However, the important point is that even in the limit $\pi\rho \ra \infty$ we are able to normalize the zero modes so that their kinetic terms are canonical in the effective 4d theory.
  
Similarly as for the gravitino, both the even and the odd components of hyperino give rise to massive KK modes on the half-circle. 
The solution is:
\bea
\phi_\lambda=e^{\frac{5}{2}b_0 k |x^5|}
( A_m J_2(\frac{m}{k}e^{b_0 k|x^5|}) +B_m Y_2(\frac{m}{k} e^{b_0 k|x^5|}))
\nn
\phi_\lambda^-=i \epsilon(x^5)e^{\frac{5}{2}b_0 k |x^5|}(
 A_m J_1(\frac{m}{k}e^{b_0 k|x^5|}) +B_m Y_1(\frac{m}{k} e^{b_0 k|x^5|}))
\eea
with the coefficients $A$, $B$ given by (\ref{ab}) and the mass $m$ given by (\ref{massq}).
All that is left is to consider the hypermultiplet scalars. Only the even scalars have massless zero modes given by:
\beq
\ln V=\sigma= const
\eeq
while the solution for massive modes is:
\bea
\ln V(x^\mu, x^5)  = \phi_V(x^5) \ln V(x^\mu) 
&
\sigma(x^\mu, x^5)  = \phi_\sigma(x^5) \sigma(x^\mu)
\nn
\xi(x^\mu,x^5)  = \phi_\xi^-(x^5) \xi(x^\mu) . 
&
{}
\eea
\bea
\phi_V=\phi_\sigma=e^{2b_0 k |x^5|}
( A_m J_2(\frac{m}{k}e^{b_0 k|x^5|}) +B_m Y_2(\frac{m}{k} e^{b_0 k|x^5|}))
\nn
\phi_\xi^-=\epsilon(x^5)e^{2b_0 k |x^5|}(
 A_m J_1(\frac{m}{k}e^{b_0 k|x^5|}) +B_m Y_1(\frac{m}{k} e^{b_0 k|x^5|}))
\eea
The multiplicative constant $A_m$ which is left undetermined in the solutions 
can be fixed by canonical normalization of the fields in the 4d effective theory. The massive modes of the scalars pair with the components of the hyperino to form  spin spectra of massive N=1 chiral multiplets.
To summarize the discussion of the KK modes we point out two   
features of the spectrum:
i) Half of the supercharges is spontaneously broken by the Randall-Sundrum vacuum solution leaving only 4 supercharges corresponding to N=1 SUSY in the 4d effective theory. And indeed, the spin spectrum of N=1 SUSY is visible at the  massless level as well as at at the level of massive KK modes. 
ii) It was found \cite{hewett} that if we place the observable sector on the TeV brane (at $x^5=\pi\rho$), the coupling of the massive modes of the graviton to the energy momentum tensor will be only TeV suppressed (compared to $M_{PL}$ suppression of the ordinary, massless graviton). In the supersymmetric extension of the Randall-Sundrum model the massive modes of the remaining bulk fields will also acquire TeV suppressed couplings to the observable sector. More precisely, if a dimension $d$ operator $\delta(x^5-x^{5i})O_d(\psi,\chi)$ is 
present in the 5d Lagrangian, where $\psi$ denotes collectively bulk fields and $\chi$ the 4d observable sector localized on a brane at $x^{5i}$, then in the effective 4d theory massive modes acquire couplings to the observable sector  suppressed by $\frac{1}{(M_{PL}e^{-b_0ky^i})^d}$. This means that the fields originating from the TeV brane couple to the massive modes of the bulk fields through higher dimensional operators suppressed approximately by $\frac{1}{(1Tev)^d}$. 

\vskip 0.5cm
To summarize , in the present paper we have cast the supersymmetric 
Randall-Sundrum model coupled to the universal hypermultiplet in the 
language of gauged supergravity. We have formulated the model in the smooth 
form, with an 
additional four-form field and a supersymmetry singlet scalar. 
The model admits 
the Poincare-foliated RS vacuum as a BPS state with four unbroken 
supercharges. 
In addition, we have discussed the structure of the KK-mode decomposition 
of fields present in the model.\\

\vspace{1.0cm} 
\noindent Authors thank Renata Kallosh for helpful discussions. 
Z.L. thanks Max Zucker for interesting discussions. \\  
\noindent This work has been supported by TMR program 
ERBFMRX--CT96--0090 and by RTN program HPRN-CT-2000-00152. 
Z.L. and S.P. are supported 
by the Polish Committee for Scientific Research grant 2 P03B 05216(99-2000).


\begin{thebibliography}{99}
\bibitem{rs1} L. Randall, R. Sundrum, {\it An alternative to compactification},  {\em Phys. Rev. Lett.} {\bf 83} (1999), hep-th/9906064.  
\bibitem{rs2} L. Randall, R. Sundrum, {\em A Large Mass Hierarchy from a Small Extra Dimension},  {\em Phys. Rev. Lett.} {\bf 83} (1999) hep-ph/9905221.
\bibitem{rs3} J. Lykken, L. Randall, {\it JHEP} {\bf 0006} (2000) 014, 
hep-th/9908076.
\bibitem{bagger} R. Altendorfer, J. Bagger and D. Nemeschansky, {\it Supersymmetric Randall-Sundrum scenario}, hep-th/0003117.  
\bibitem{gp} T. Gherghetta, A. Pomarol, {\it Bulk fields and supersymmetry 
in a slice of AdS}, hep-th/0003129.
\bibitem{hiszp} N. Alonso-Alberca, P. Meesen, T. Ortin, {\it Supersymmetric  
Brane-Worlds}, hep-th/0003248.  
\bibitem{flp} A. Falkowski, Z. Lalak and S. Pokorski, {\it Supersymmetrizing branes with bulk in five-dimensional supergravity}, hep-th/0004093, {\it Phys. Lett } {\bf B} in press.   
\bibitem{mt} A. Falkowski, {\it M.Sc.Thesis}, info.fuw.edu.pl/\~{}afalkows.
\bibitem{kallosh} E. Bergshoeff, R. Kallosh and A. Van Proeyen, {\it Supersymmetry in singular spaces}, hep-th/0007044.
\bibitem{bc} K. Behrndt, M. Cveti\'{c}, {\it Phys. Rev.} {\bf D61} (2000)  
101901.  
\bibitem{bc2} K. Behrndt, M. Cveti\'{c}, {\it Phys. Lett.} {\bf B475} (2000) 
253. 
\bibitem{kl} R. Kallosh, A. Linde, {\it JHEP} {\bf 0002} (2000) 0005 and references therein.   
\bibitem{kl2} R. Kallosh, A. Linde and M. Shmakova, {\it JHEP} {\bf 9911} (1999) 010.
\bibitem{kb} K. Behrndt, {\it Nonsingular infrared flow from D=5 gauged supergravity}, hep-th/0005185.
\bibitem{pmayr} P. Mayr, {\it Stringy world branes and exponential hierarchies}, hep-th/0006204. 
\bibitem{duff} M. J. Duff, J. T. Liu, K. S. Stelle, {\it A supersymmetric Type IIB Randall-Sundrum realization}, hep-th/0007120. 
\bibitem{louis} K. Behrndt, C. Herrmann, J. Louis, and S. Thomas, 
{ \it Domain walls in five dimensional supergravity with non-trivial hypermultiplets}, hep-th/0008112.
\bibitem{zucker} M. Zucker, {\em Supersymmetric Brane Worlds Scenarios 
from Off-Shell Supergravity}, hep-th/0009083.
\bibitem{gunaydin} M. Gunaydin, G. Sierra and  P. K. Townsend, {\it Nucl.Phys.}{\bf B253} (1985) 573;\\
M. Gunaydin and M. Zagermann, {\it Nucl.Phys.} {\bf B572} (2000), 
 hep-th/9912027.     
\bibitem{agata} A. Ceresole, G. Dall'Agata, {\it General matter coupled ${\cal N}=2$, D=5 gauged supergravity}, hep-th/0004111.
\bibitem{ovrutdw} A. Lukas,
B. A. Ovrut, K. S. Stelle and D. Waldram, {\it The universe as a domain wall},
{\em Phys. Rev.} {\bf D59} (1999) 086001, hep-th/9806051.
\bibitem{ovruthet} A. Lukas, B. A. Ovrut, K.S. Stelle and D. Waldram, {\it Heterotic M-theory in Five Dimensions} {\em Nucl. Phys.} {\bf B552} (1999) 246,
hep-th/9806051.
\bibitem{elp} J. Ellis, Z. Lalak, W. Pokorski,  {\em Five-Dimensional Gauged Supergravity and Supersymmetry Breaking in M Theory}, {\em Nucl. Phys.} {\bf B559} (1999) 71, hep-th/9811133. 
\bibitem{fs} S. Ferrara and S. Sabharwal, {\em Nucl. Phys.} {\bf B332} (1990) 317.
\bibitem{hewett}
 H. Davoudiasl, J.L. Hewett and T.G. Rizzo, {\it Phenomenology of the Randall-Sundrum gauge hierarchy}, {\em Phys. Rev. Lett.} {\bf 84} (2000) 2080, hep-ph/9909255.  

\end{thebibliography}
\end{document}